# Ensemble Asteroseismology of Solar-Type Stars with the NASA Kepler Mission

W. J. Chaplin[*,1], H. Kjeldsen[2], J. Christensen-Dalsgaard[2], S. Basu[3], A. Miglio[1,4],
T. Appourchaux[5], T. R. Bedding[6], Y. Elsworth[1], R. A. García[7], R. L. Gilliland[8], L. Girardi[9],
G. Houdek[10], C. Karoff[2], S. D. Kawaler[11], T. S. Metcalfe[12], J. Molenda-Żakowicz[13],
M. J. P. F. G. Monteiro[14], M. J. Thompson[12], G. A. Verner[15,1], J. Ballot[16], A. Bonanno[17],
I. M. Brandão[14], A.-M. Broomhall[1], H. Bruntt[2], T. L. Campante[14,2], E. Corsaro[17],
O. L. Creevey[18,19], G. Doğan[2], L. Esch[3], N. Gai[21,3], P. Gaulme[5], S. J. Hale[1], R. Handberg[2],
S. Hekker[21,1], D. Huber[6], A. Jiménez[18,19], S. Mathur[12], A. Mazumdar[22], B. Mosser[23], R. New[24],
M. H. Pinsonneault[25], D. Pricopi[26], P.-O. Quirion[27], C. Régulo[18,19], D. Salabert[18,19],
A. M. Serenelli[28], V. Silva Aguirre[29], S. G. Sousa[14], D. Stello[6], I. R. Stevens[1], M. D. Suran[26],
K. Uytterhoeven[7], T. R. White[6], W. J. Borucki[30], T. M. Brown[31], J. M. Jenkins[32],
K. Kinemuchi[33], J. Van Cleve[32], T. C. Klaus[34]

1 School of Physics and Astronomy, University of Birmingham, Edgbaston, Birmingham, B15 2TT, UK

2 Department of Physics and Astronomy, Aarhus University, DK-8000 Aarhus C, Denmark

3 Department of Astronomy, Yale University, P.O. Box 208101, New Haven, CT 06520-8101, USA

4 Département d'Astrophysique, Géophysique et Océanographie (AGO), Université de Liège, Allée du 6 Août 17 4000 Liège 1, Belgique

5 Institut d'Astrophysique Spatiale, Université Paris XI−CNRS (UMR8617), Batiment 121, 91405 Orsay Cedex, France

6 Sydney Institute for Astronomy (SIfA), School of Physics, University of Sydney, NSW 2006, Australia

7 Laboratoire AIM, CEA/DSM−CNRS−Université Paris Diderot ; IRFU/SAp, Centre de Saclay, 91191 Gif-sur-Yvette Cedex, France

8 Space Telescope Science Institute, Baltimore, MD 21218, USA

9 Osservatorio Astronomico di Padova, INAF, Vicolo dell'Osservatorio 5, I-35122 Padova, Italy

10 Institute of Astronomy, University of Vienna, A-1180 Vienna, Austria

11 Department of Physics and Astronomy, Iowa State University, Ames, IA 50011, USA

12 High Altitude Observatory and, Scientific Computing Division, National Center for Atmospheric Research, Boulder, Colorado 80307, USA

13 Astronomical Institute, University of Wrocław, ul. Kopernika, 11, 51-622 Wrocław, Poland

14 Centro de Astrofísica and Faculdade de Ciências, Universidade do Porto, Rua das Estrelas, 4150-762, Portugal

15 Astronomy Unit, Queen Mary, University of London, Mile End Road, London, E1 4NS, UK

16 Institut de Recherche en Astrophysique et Planétologie, Université de Toulouse, CNRS, 14 av E. Belin, 31400 Toulouse, France

17 INAF Osservatorio Astrofisico di Catania, Via S.Sofia 78, 95123, Catania, Italy

18 Departamento de Astrofísica, Universidad de La Laguna, E-38206 La Laguna, Tenerife, Spain

19 Instituto de Astrofísica de Canarias, E-38200 La Laguna, Tenerife, Spain

20 Department of Physics, Dezhou University, Dezhou 253023, China

21 Astronomical Institute, "Anton Pannekoek", University of Amsterdam, PO Box 94249, 1090 GE Amsterdam, The Netherlands

22 Homi Bhabha Centre for Science Education (TIFR), V. N. Purav Marg, Mumbai 400088, India

23 LESIA, CNRS, Université Pierre et Marie Curie, Université Denis Diderot, Observatoire de Paris, 92195 Meudon cedex, France

24 Materials Engineering Research Institute, Faculty of Arts, Computing, Engineering and Sciences, Sheffield Hallam University, Sheffield, S1 1WB, UK

25 Department of Astronomy, The Ohio State University, 4055 McPherson Laboratory, 140 West 18th Avenue, Columbus, OH 43210, USA

26 Astronomical Institute of the Romanian Academy, Str. Cutitul de Argint, 5, RO 40557, Bucharest, RO

27 Canadian Space Agency, 6767 Boulevard de l'Aéroport, Saint-Hubert, QC, J3Y 8Y9, Canada

28 Instituto de Ciencias del Espacio (CSIC-IEEC), Campus UAB, Facultad de Cìencies, Torre-C5, 08193, Bellaterra, Spain

29 Max Planck Institute for Astrophysics, Karl Schwarzschild Str. 1, Garching, D-85741, Germany

30 NASA Ames Research Center, MS 244-30, Moffett Field, CA 94035, USA

31 Las Cumbres Observatory Global Telescope, Goleta, CA 93117, USA

32 SETI Institute/NASA Ames Research Center, Moffett Field, CA 94035, USA

33 Bay Area Environmental Research Institute/NASA Ames Research Center, Moffett Field, CA 94035, USA

34 Orbital Sciences Corporation/NASA Ames Research Center, Moffett Field, CA 94035, USA

*To whom correspondance should be addressed. E-mail w.j.chaplin@bham.ac.uk

**In addition to its search for extra-solar planets, the NASA Kepler Mission provides exquisite data on stellar oscillations. We report the detections of oscillations in 500 solar-type stars in the Kepler field of view, an ensemble that is large enough to allow statistical studies of intrinsic stellar properties (such as mass, radius and age) and to test theories of stellar evolution. We find that the distribution of observed masses of these stars shows intriguing differences to predictions from models of synthetic stellar populations in the Galaxy.**

An understanding of stars is of central importance to astrophysics. Uncertainties in stellar physics have a direct impact on fixing the ages of the oldest stellar populations (which place tight constraints on cosmologies) as well as on tracing the chemical evolution of galaxies. Stellar astrophysics also plays a crucial role in the current endeavors to detect habitable planets around other stars (*1 - 5*). Accurate data on the host stars are required to determine the sizes of planets discovered by the transit method, to fix the locations of habitable zones around the stars, and to estimate the ages and understand the dynamical histories of these stellar systems. Measurements of the levels of stellar activity, and their variations over time (*6*), provide insights into planetary habitability, the completeness of the survey for extrasolar planets, and on the surface variability shown by our own Sun, which has very recently been in a quiescent state that is unique in the modern satellite era (*7,8*).

New insights are being made possible by asteroseismology, the study of stars by observations of their natural, resonant oscillations (*9, 10*). Stellar oscillations are the visible manifestations of standing waves in the stellar interiors. Main-sequence and sub-giant stars whose outer layers are unstable to convection (solar-type stars) display solar-like oscillations that are predominantly acoustic in nature, excited by turbulence in the convective envelopes (*11, 12*). The dominant oscillation periods are minutes in length and give rise to variations in stellar brightness at levels of typically just a few parts per million. The frequencies of the oscillations depend on the internal structures of the stars and their rich information content means that the fundamental stellar properties (e.g., mass, radius, and age) can be determined to levels that are difficult to achieve by other means, while the internal structure and dynamics can be investigated in a unique way.

Helioseismology has provided us with an extremely detailed picture of the internal structure and dynamics of the Sun, including tests of basic physics (*13 - 15*). Such investigations are beginning to be possible for other stars. Over the last decade the quality of seismic observations on other solar-type stars has been improving steadily, from ground-based spectroscopy (*16 - 18*) and the French-led CoRoT (Convection Rotation and Planetary Transits) satellite (*19, 20*). Now, Kepler is providing ultra-precise observations of variations in stellar brightness (photometry), which are suitable for the study of solar-like oscillations (*21*). During the first 7 months of science operations more than 2000 stars were selected for observation for 1 month each with a cadence rapid enough to perform an asteroseismic survey of the solar-type population in the Kepler field of view. Here, we report the detection of solar-like oscillations in 500 of those stars. Previously, this type of oscillation had been detected in only about 25 stars.

As is evident from the frequency spectra of the oscillations exhibited by nine stars from the ensemble (Fig. 1), solar-like oscillators present a rich, near-regular pattern of peaks that are the signatures of high-order overtones. The dominant frequency spacing is the so-called large separation, $\Delta\nu$, between consecutive overtones (*22*). The average large separation scales approximately with the square root of the mean density of the star. The observed power in the oscillations is modulated in frequency by a Gaussian-like envelope. The frequency of maximum oscillation power, $\nu_{max}$, scales approximately as $gT_{eff}^{-1/2}$, where $g \propto M/R^2$ is the surface gravity and $T_{eff}$ is the effective temperature of the star (*23, 24*).

Figure 2 shows all the stars on a conventional Hertzsprung-Russell diagram, which plots the luminosities of stars against $T_{eff}$. The temperatures were estimated (*25*) from multicolor photometry available in the Kepler Input Catalog (*26*). Luminosities were estimated from the temperatures and the seismically estimated radii [see below and (*27*)]. We also plot $\Delta\nu$ against temperature and, just like the conventional diagram, this asteroseismic version delineates different types of stars and different evolutionary states (the $\nu_{max}$ version is similar). Main-sequence stars, burning hydrogen into helium in their cores, lie in a diagonal swathe (from the lower right to top left) on each diagram. Both asteroseismic parameters, $\Delta\nu$ and $\nu_{max}$, decrease along the main sequence toward hotter solar-type stars, where surface gravities and mean densities are lower than in cooler stars (and luminosities are higher). After exhaustion of the core hydrogen, stars eventually follow nearly horizontal paths in the luminosity plot toward lower temperatures as they evolve as sub-giants, before turning sharply upwards to become red giants (*28, 29*). The values of $\Delta\nu$ and $\nu_{max}$ decrease comparatively rapidly through the sub-giant phase. Detailed information on the physics of the interiors of these stars is emerging from analysis of Kepler data (*30*).

We have detected solar-like oscillations in relatively few stars that have $\Delta\nu$ and $\nu_{max}$ larger than the solar values. These stars are intrinsically fainter, and less massive, than the Sun, and we see fewer detections because the intrinsic oscillation amplitudes are lower than in the hotter main-sequence and evolved sub-giant stars. This detection bias means that the most populous cohort in the ensemble is that comprising sub-giants. Sub-giants have more complicated oscillation spectra than main-sequence stars. The details of the spectra depend on how, for example, various elements are mixed both within and between different layers inside the stars. Seismic analysis of the Sun has already shown that merely reproducing the luminosity and temperature of a star will not guarantee that the internal structure, and hence the underlying physics, is correct. This inspired the inclusion of additional physics, such as the settling over time of chemical elements due to gravity, in stellar models (*13*). The Sun is a relatively simple star compared to some of the solar-type stars observed by Kepler.

We have made use of the $\Delta\nu$ and $\nu_{max}$ of the ensemble together with photometric estimates of the temperatures to estimate the masses and radii of the stars in a way that is independent of stellar evolutionary models—using the so-called direct-method of estimation (*27*)—and then compared the observed distributions with those predicted from synthetic stellar populations (Fig. 3). The synthetic populations were calculated by modelling the formation and evolution of stars in the Kepler field of view, which lies in the Cygnus region of the Orion arm of our galaxy, the Milky Way (*27*). This modeling requires descriptions of, for example, the star formation history (including the frequency of occurrence of stars with various masses), the spatial density of stars in the disc of the Milky Way, and the rate at which the galaxy is chemically enriched by stellar evolution (*31*).

Previous population studies have been hampered by not having robust mass estimates on individual stars (*31*). Precise estimates of masses of solar-type stars had been limited principally to stars in eclipsing binaries (*32*). The Kepler estimates add substantially to this total, and in numbers that are large enough to do statistical population tests using direct mass estimates, which has not been possible before.

While the distributions of stellar radii in Fig. 3 are similar, the same cannot be said for the mass distributions. We have quantified the significance of the differences using statistical tests. Differences in radius were judged to be marginally significant at best. In contrast those in mass were found to be highly significant (>99.99%) (*27*). The observed distribution of masses is wider at its peak than the modeled distribution, and is offset towards slightly lower masses.

Tests suggest that, for the bulk of the stars, bias in the estimated masses and radii is no larger than the estimated uncertainties (*27*). On the assumption that the observed masses and radii are robust, this result may have implications for both the star formation rate and the initial mass function of stars. Mixing or overshooting of material between different layers (including stellar cores), and the choice of the so-called mixing length parameter, which measures the typical lengthscale of the convection and is one of the few free parameters in stellar evolution theory, may also be relevant. It is yet to be tested whether the expected small fraction of unresolved binaries could have contributed to the mass discrepancy.

36. Kepler is a NASA Discovery Class Mission, which was launched in March 2009, whose funding is provided by NASA's Science Mission Directorate. The authors thank the entire Kepler team, without whom these results would not be possible. The asteroseismology program of Kepler is being conducted by the Kepler Asteroseismology Science Consortium.

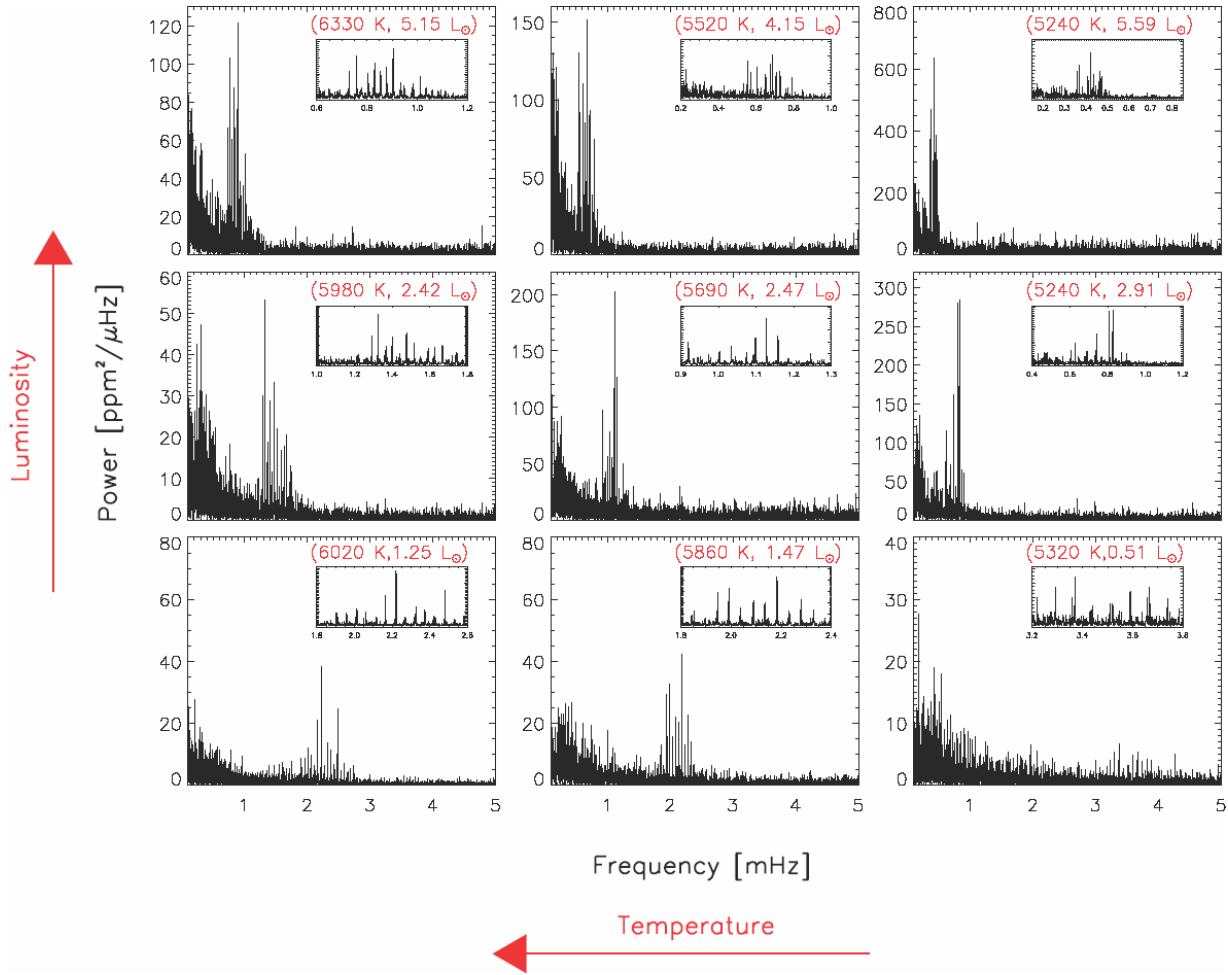

**Fig 1.** Frequency spectra of the oscillations exhibited by nine stars from the ensemble. Each spectrum shows a prominent Gaussian-shaped excess of power due to the oscillations, centered on the frequency $\nu_{max}$. (**Insets**) Clearer views of the near-regular spacings in frequency between individual modes of oscillation within each spectrum. The stars are arranged by intrinsic brightness [in units of luminosity ($L_\odot$)] and temperature, with intrinsically fainter stars showing weaker, less prominent oscillations than their intrinsically brighter cousins. ppm, parts per million.

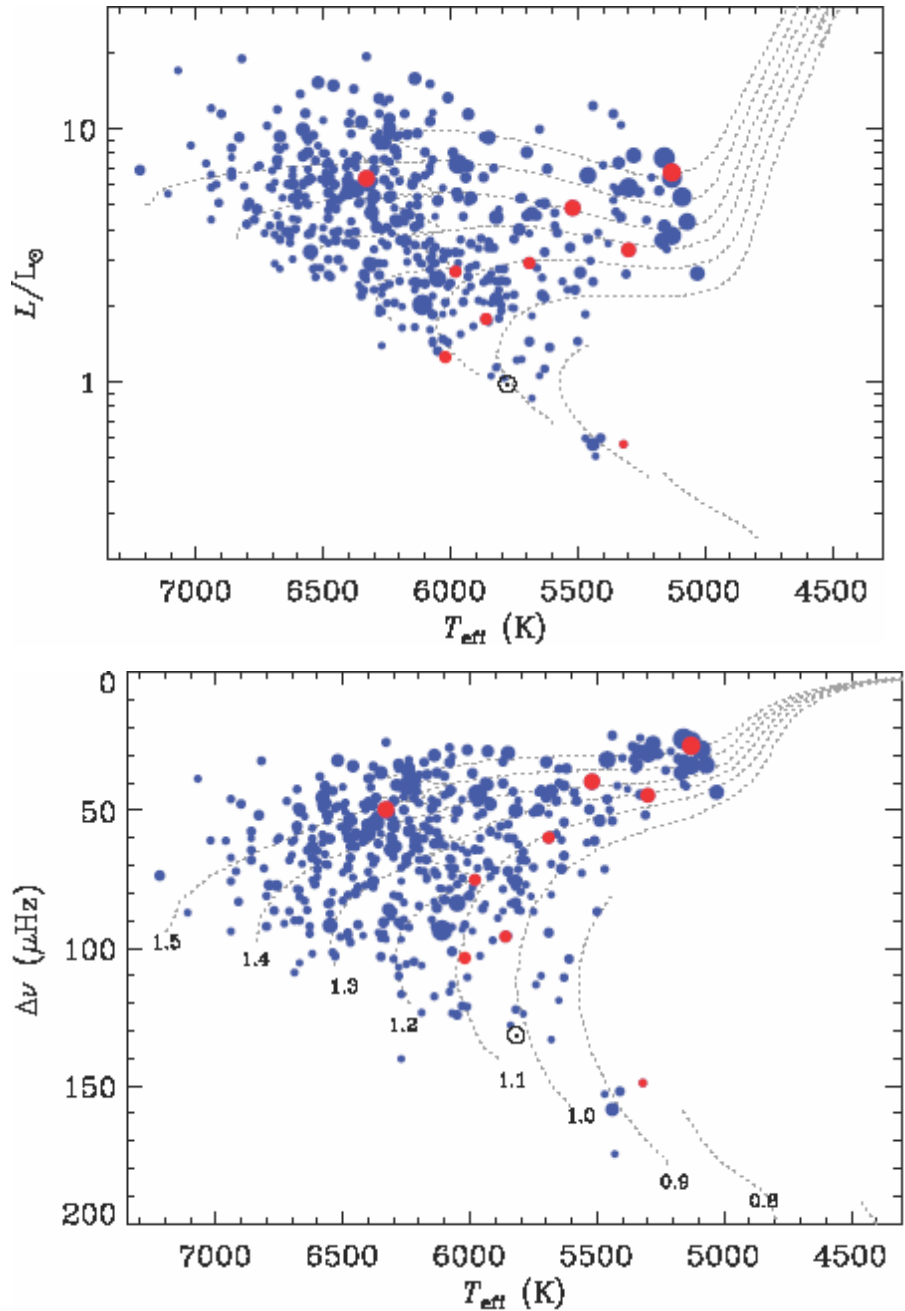

**Fig 2.** (**Top**) Estimates of the luminosities of the stars (in units of the solar luminosity) of the ensemble of Kepler stars showing detected solar-like oscillations, plotted as a function of effective temperature. Stars from Fig. 1 are plotted with red symbols. (**Bottom**) Average large frequency separations, $\Delta\nu$, against effective temperature. The symbol sizes are directly proportional to the prominence of the detected oscillations (i.e., the signal-to-noise ratios). These ratios depend both on stellar properties (e.g. the photometric amplitudes shown by the oscillations, and the intrinsic stellar backgrounds from convection) and the apparent brightness of the stars. The dotted lines show predicted evolutionary tracks (*33*) for models of different stellar mass (0.8 to 1.5 solar masses, in steps of 0.1). The Sun is marked with a solar symbol ($\odot$).

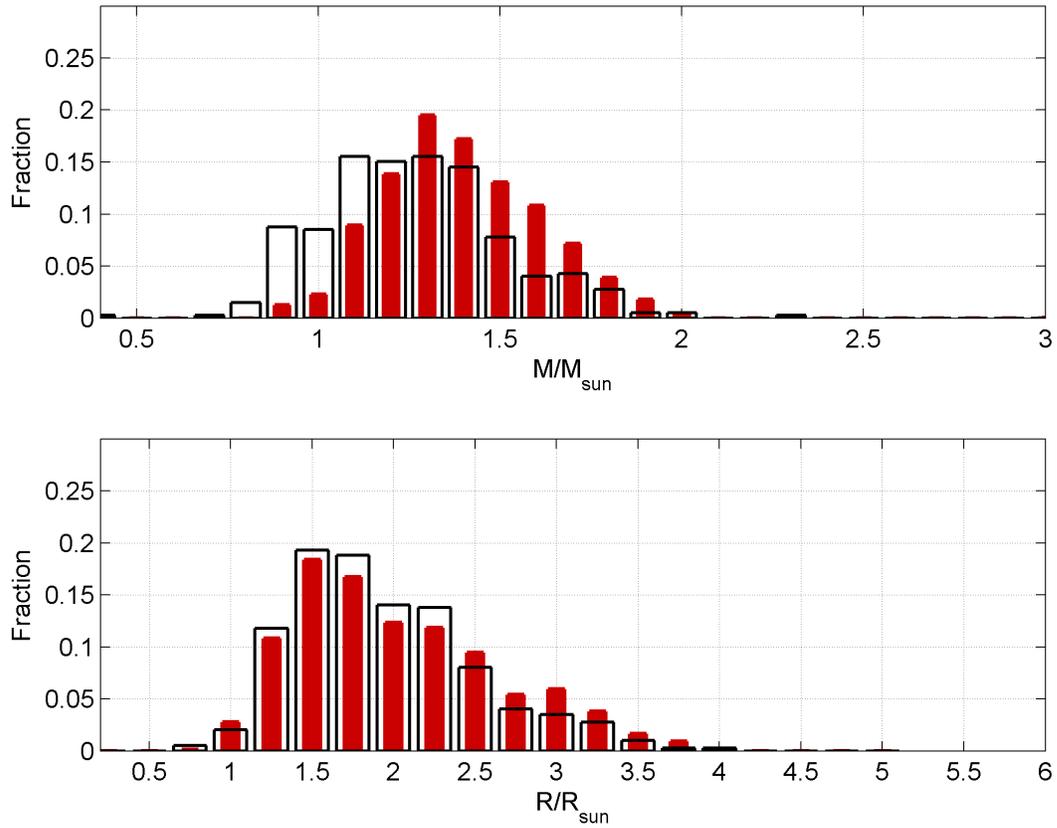

**Fig. 3.** Black lines: Histograms of the observed distribution of masses (**top**) and radii (**bottom**) of the *Kepler* ensemble (*27*). In red, the predicted distributions from population synthesis modelling, after correction for the effects of detection bias (*27*). The population modeling was performed using the TRIGEGAL code (*34, 35*).

# Online Material

**Ensemble asteroseismology of solar-type stars with the NASA Kepler Mission**

## 1. Observations

The primary objective of the NASA Kepler Mission is to detect, by the transit method, Earth-sized planets in the habitable zones of solar-type stars. Photometry of most of the stars is conducted at a long cadence (LC) of 29.4 minutes, but a subset of up to 512 stars can be observed at any one time at a short cadence (SC) of 58.85 s (S1, S2). The cadence of the SC data is rapid enough to allow investigations of solar-like oscillations in main-sequence stars, where dominant periods are of the order of several minutes.

We use asteroseismic results on solar-type stars that were observed by Kepler during the first seven months of science operations. About 2000 stars, down to Kepler apparent magnitude $Kp \approx 12$, were selected as potential solar-type targets based upon parameters in the Kepler Input Catalog (S3, S4). Each star was observed for one month at a time in SC mode. Time series were prepared for asteroseismic analysis using procedures that work on the raw lightcurves which were developed for application to GOLF/SoHO data (S5). Additive corrections were applied to correct thermal drifts; any sudden jumps or discontinuities were removed; while outliers were removed by clipping at the $5\sigma$ level.

## 2. Estimation of average seismic parameters

The frequency power spectra of "solar-like" oscillations in solar-type stars present a pattern of peaks with near regular frequency separations (see Figs S2 and S3). The mode powers are modulated by an envelope that has a bell-shaped (i.e., Gaussian) appearance in many stars for which solar-like oscillations have been observed (including the Sun). Different techniques have been devised and applied to the Kepler SC data to detect signatures of solar-like oscillations (e.g., S6, S7, S8, S9, S10, S11, S12 and S13). Some techniques rely on extracting signatures of the near-regular frequency separations of the oscillations, while others search for signatures of the Gaussian-like power excess due to the oscillations. These methods have been tested intensively on artificial datasets (e.g., as part of the asteroFLAG hare-and-hounds exercises; see S14, S15). The primary data products from these automated searches are estimates of the average large frequency separation, $\Delta\nu$, and the frequency of maximum oscillations power, $\nu_{max}$ (see Fig. S1; see also Fig. 2 in main article).

The large frequency separation, $\Delta\nu$, is the spacing between consecutive overtones of the same spherical angular degree, $l$. When the signal-to-noise ratios in the seismic data are insufficient to allow robust extraction of individual oscillation frequencies, it is still possible to extract estimates of the average large frequency separation for use as the seismic input data. Indeed, this is the case for many Kepler stars. The average large separation scales to good approximation with the square root of the mean density of a star (S16). This gives the following scaling relation:

$$\left(\frac{\Delta\nu}{\Delta\nu_{Sun}}\right) \approx \left(\frac{M}{M_{Sun}}\right)^{0.5}\left(\frac{R}{R_{Sun}}\right)^{-1.5}, \qquad (1)$$

with $M$ and $R$ the stellar mass and radius, and $\Delta\nu_{Sun}$ the large separation of the Sun.

The frequency of maximum power in the oscillations power spectrum, $\nu_{max}$, is related to the acoustic cut-off frequency of a star (S17, S18, S19, S20), which in turn scales to very good approximation as $MR^{-2} T_{eff}^{-0.5}$ with $T_{eff}$ the effective temperature of the star. This gives the following scaling relation:

$$\left(\frac{\nu_{max}}{\nu_{max,Sun}}\right) \approx \left(\frac{M}{M_{Sun}}\right)\left(\frac{R}{R_{Sun}}\right)^{-2}\left(\frac{T_{eff}}{T_{eff,Sun}}\right)^{-0.5}. \qquad (2)$$

3. Estimation of $M$ and $R$

3.1. Direct method of estimation

If $\Delta\nu$, $\nu_{max}$ and $T_{eff}$ are known, Equations (1) and (2) represent two equations in two unknowns, and we may therefore re-arrange and solve for $M$ and $R$ in what we call the "direct" method of estimation of the stellar properties. This gives (S21):

$$\left(\frac{R}{R_{Sun}}\right) = \left(\frac{\nu_{max}}{\nu_{max,Sun}}\right)\left(\frac{\Delta\nu}{\Delta\nu_{Sun}}\right)^{-2}\left(\frac{T_{eff}}{T_{eff,Sun}}\right)^{0.5}, \qquad (3)$$

and

$$\left(\frac{M}{M_{Sun}}\right) = \left(\frac{\nu_{max}}{\nu_{max,Sun}}\right)^3 \left(\frac{\Delta\nu}{\Delta\nu_{Sun}}\right)^{-4} \left(\frac{T_{eff}}{T_{eff,Sun}}\right)^{1.5}. \quad (4)$$

The results presented in the paper were estimated by the direct method. We used $\nu_{max,Sun}$=3150 µHz and $\Delta\nu_{Sun}$=134.9 µHz. Estimates of $T_{eff}$ were derived from the multicolor photometry available in the Kepler Input Catalog (based on the approach discussed in S22). The median fractional precision in the temperatures is about 1 %.

We obtained from the direct method a median fractional uncertainty of just over 10 % in $M$ and about 5.5 % in $R$. That the fractional uncertainties on $M$ are larger than those on $R$ is apparent from the propagation of the uncertainties on the observables, i.e.

$$\left(\frac{\delta R}{R}\right)^2 = \left(\frac{\delta\nu_{max}}{\nu_{max}}\right)^2 + \left(2\frac{\delta\Delta\nu}{\Delta\nu}\right)^2 + \left(0.5\frac{\delta T_{eff}}{T_{eff}}\right)^2, \quad (5)$$

and

$$\left(\frac{\delta M}{M}\right)^2 = \left(3\frac{\delta\nu_{max}}{\nu_{max}}\right)^2 + \left(4\frac{\delta\Delta\nu}{\Delta\nu}\right)^2 + \left(1.5\frac{\delta T_{eff}}{T_{eff}}\right)^2. \quad (6)$$

The direct method is very attractive because it provides mass and radius estimates that are independent of stellar evolutionary models, and this is clearly of great benefit for instructive comparisons with the population synthesis models. However, the direct method does give larger uncertainties on $M$ and on $R$ than would be obtained from a "grid-based" method of estimation (here about twice as large; see Section 3.2 below). Although the uncertainties are larger than for the grid-based method, this does mean that they are expected to largely capture any uncertainties in Equations (1) and (2) due to, for example, metallicity effects.

For the present, the lack of precise independent constraints on the metallicities (e.g., on [Fe/H]) means that the grid method is vulnerable to systematic bias in the estimates of $M$ (although not $R$). Once we have those tight independent constraints on all the stars – from complementary ground-based observations being made in support of Kepler – we will be able to take full advantage of the grid-based method and the significantly better precision it offers, as we now go on to discuss in Section 3.2. After that, we shall return in Section 3.3 to discuss levels of systematic bias in the direct method.

## 3.2. Grid-based method of estimation

We also applied the so-called grid-based method to estimation of the $M$ and $R$ of the Kepler stars. This is essentially the well-used approach in stellar astronomy of matching the observations to stellar evolutionary tracks, but with the powerful diagnostic information contained in the seismic $\Delta\nu$ and $\nu_{max}$ also brought to bear (e.g., as per Equations (1) and (2)). Properties of stars are determined by searching among a grid of stellar evolutionary models to get a "best fit" for a given observed set of input parameters, typically $\{\Delta\nu, \nu_{max}, T_{eff}, [Fe/H]\}$. While the direct method assumes that all values of temperature are possible for a star of a given mass and radius, we know from stellar evolution theory that only a narrow range of $T_{eff}$ is allowed for a given $M$ and $R$. This prior information is implicit in the grid-based approach, and means that estimated uncertainties on $M$ and $R$ are lower than for the direct method because a narrower range of outcomes is permitted.

Descriptions of the various grid-based pipelines used in the analysis of Kepler data may be found in, for example, S15, S23, S24, and S25. In addition to $\Delta\nu$, $\nu_{max}$ and $T_{eff}$, the grid-based methods also used as input the [Fe/H] provided in the Kepler Input Catalog. The KIC [Fe/H] are derived from the Kepler photometry, but since the available wavelength bands are not optimally sensitive to metallicity the uncertainties are very large (taken to be 0.5 dex).

The grid-based estimates of $M$ are sensitive to choices made in construction of the grid of stellar models, i.e., they are to some extent model dependent (see S25 for an in-depth discussion). The biggest sensitivity is to the metallicity. One might have thought that the large uncertainties assumed on the KIC [Fe/H] would have largely captured the range of possibilities, albeit at the cost of reducing the precision in the estimated $M$ (even then, we obtained a typical fractional median error on $M$ of about 6 %). However, without tight prior constraints on metallicity, the estimated masses are still affected by the helium abundance, $Y$, of models in the grid (often treated in different ways). This is why it is extremely important to have precise independent constraints on [Fe/H] for all the stars.

Inconsistencies resulting from the poorly known metallicity and the treatment of Y are then removed, and it will be possible to take full advantage of the superior precision in M offered by the grid-based approach.

As shown by the extensive tests of the method (S15, S25) the grid-based estimates of R are, in contrast, largely insensitive to the choice of grid, i.e., they may be regarded as being essentially model independent, at the typical level of precision given by the one-month Kepler survey data. We found that estimates of R for the Kepler stars returned by the direct and grid-based methods were indeed consistent.

### 3.3. Systematic bias in direct method of estimation

The accuracy of Equations (1) and (2) will determine levels of systematic bias in masses and radii estimated by the direct method.

The accuracy of Equation (1) may be tested using stellar evolutionary models, by comparing the average $\Delta\nu$ implied by the M and R of each model with the average $\Delta\nu$ of the model's computed oscillation frequencies. Comparisons of this type suggest that for the bulk of stars here – which cover the effective temperature range 5000 to 6750K and are on average slightly metal poor compared to the Sun – Equation (1) is accurate to 2 to 3%. We note that such a comparison does not allow for the impact of the so-called "surface term" (S26) on the observed average $\Delta\nu$. However, at least in so far as the Sun is concerned, that effect is small (e.g., see the discussion in S25, which suggests that any bias is at less than the 1% level, and within the observational uncertainties on $\Delta\nu$ obtained in this paper).

Testing the accuracy of the equation for $\nu_{max}$ (Equation (2)) is less straightforward using models, because we have much less confidence in theoretical predictions of the excitation and damping rates of the modes – which are needed to make model predictions of $\nu_{max}$ – than we do in theoretical predictions of the oscillation frequencies. However, we can instead gain some insight from results on the real Kepler data.

First, we use the grid-based approach to estimate the surface gravity, $g$, of each star, using the average $\Delta\nu$ (plus $T_{eff}$ and [Fe/H]) as input, but not $\nu_{max}$. As per the radius, R, this returns robust estimates of $g$, which are largely insensitive to the metallicity and the choice of model grid. We then use Equation (2) to estimate $g$ "directly", using $\nu_{max}$ and $T_{eff}$ only. The fractional difference in the two estimates of $g$ contains information on the bias in $\nu_{max}$. From solar temperature up to 6500K, the $g$ are found to agree to within 2.5%. The agreement is not as good for the coolest and hottest stars in the ensemble, reaching ≈ 5% at both 5000K and 6750K. The results here suggest that the $\nu_{max}$ scaling is probably good to a few per cent.

The combined upshot of the above is to suggest net bias on estimates of mass and radius that for most of the stars in the Kepler ensemble is at a level that is probably no larger than the quoted uncertainties on $M$ and $R$ (given in Section 3.1 above). The bias may be larger than this at the hottest and coolest temperatures.

We may also check the scalings using asteroseismic results on $\Delta\nu$ and $\nu_{max}$ obtained from Doppler velocity data on very bright solar-type stars, where strong constraints on the stellar properties are available from non-seismic data. An excellent, up-to-date compendium of available results is presented for 23 stars in S27. Three bright, well-studied stars (Procyon A and α Cen A and B) have precise estimates of $M$ from solutions to their binary orbits. Masses estimated using the direct method agree with the binary-solution masses to within a few per cent (within the observational uncertainties). Many of the stars in S27 have precise estimates of radii, $R$, from interferometry; precise parallaxes on the stars also yield precise estimates of $R$. Comparison of the direct-method estimates of $R$ with the interferometric and parallax-estimated $R$ again yields good agreement.

## 4. Population synthesis modeling

We estimated the properties of the stellar population observed by Kepler using the code TRILEGAL (S28), designed to simulate photometric surveys in the Galaxy. In TRILEGAL, several model parameters (such as the star-formation history and the morphology of different galactic components) were calibrated to fit Hipparcos data for the immediate solar neighborhood (S29), as well as star counts from a wide area (with 2MASS; S30), and a few deep photometric surveys, i.e., CDFS (S31), and DMS (S32). We adopted the standard parameters describing the components of the Galaxy and simulated the stellar population in the sky area observed in each of the 21 five-square-degree Kepler sub-fields of view, considering for each of them an average interstellar extinction at infinity (S33). The extinction is assumed to be caused by an exponential dust layer with a scale height, above and below the galactic plane, equal to 110 parsec. The photometry in TRILEGAL was simulated with the known wavelength response function of Kepler, and the synthetic population was magnitude-selected, using the same range as the observed sample (see Section 1 above).

We also simulated observing each of the stars in the synthetic population for one month at a time with Kepler, in order to predict whether the stars would show detectable solar-like oscillations. Those synthetic stars judged to have detectable oscillations were added to a final list for direct comparison with the observations. In brief, we used the known $M$, $R$ and $T_{\text{eff}}$ of each synthetic star to predict the total mean power we would expect to observe in its solar-like oscillations, using well-established scaling relations that describe the seismic parameters in terms of the stellar properties (e.g., S17, S18, S19). From the distance of each synthetic star, and the known luminosity, we could calculate an apparent magnitude and, from that, the noise that would be expected in observations (using the description in S1). We then applied statistical tests to the resulting S/N estimate to estimate the probability of detecting solar-like oscillations, assuming observations lasting one month. The tests were based on well-established approaches used in helioseismology and asteroseismology (S34, S35). Synthetic stars were included in the final synthetic distribution if there was judged to be a better than 9 in 10 chance of making a detection.

## 5. Comparison of observed and synthetic distributions

We applied the Kolmogorov-Smirnov (K-S) test in order to quantify differences between the observed and synthetic distributions. The K-S test returns an estimate of the probability that the observed and synthetic distributions come from the same, parent population. We applied the test in a way that took into account the statistical uncertainties on the observed masses and radii. Take, for example, the test as applied to the masses. To each observed mass we added a random offset drawn from a Gaussian distribution having a standard deviation equal to the estimated uncertainty on the mass. The K-S test was then applied to the resulting, perturbed set of observed masses. The test was repeated 5000 times, each time on a fresh set of perturbed masses (made with a new set of random numbers). The distribution of results from the 5000 K-S tests then allowed us to judge the full range of probabilities commensurate with the observational uncertainties. (We used results from artificial data to calibrate the impact of uncertainties of different fractional sizes on the returned K-S probabilities, in cases where the artificial "observations" were known to come from the same population as the artificial "synthetic" data.)

The K-S probabilities for mass were in every case lower than $10^{-5}$, indicating that differences between the observed and synthetic distributions were highly significant. K-S probabilities for radius gave values that in contrast went up to approximately 20 %, with a typical value between 5 and 10 %. These levels of probability (in what is a null hypothesis test) are only marginally significant at best.

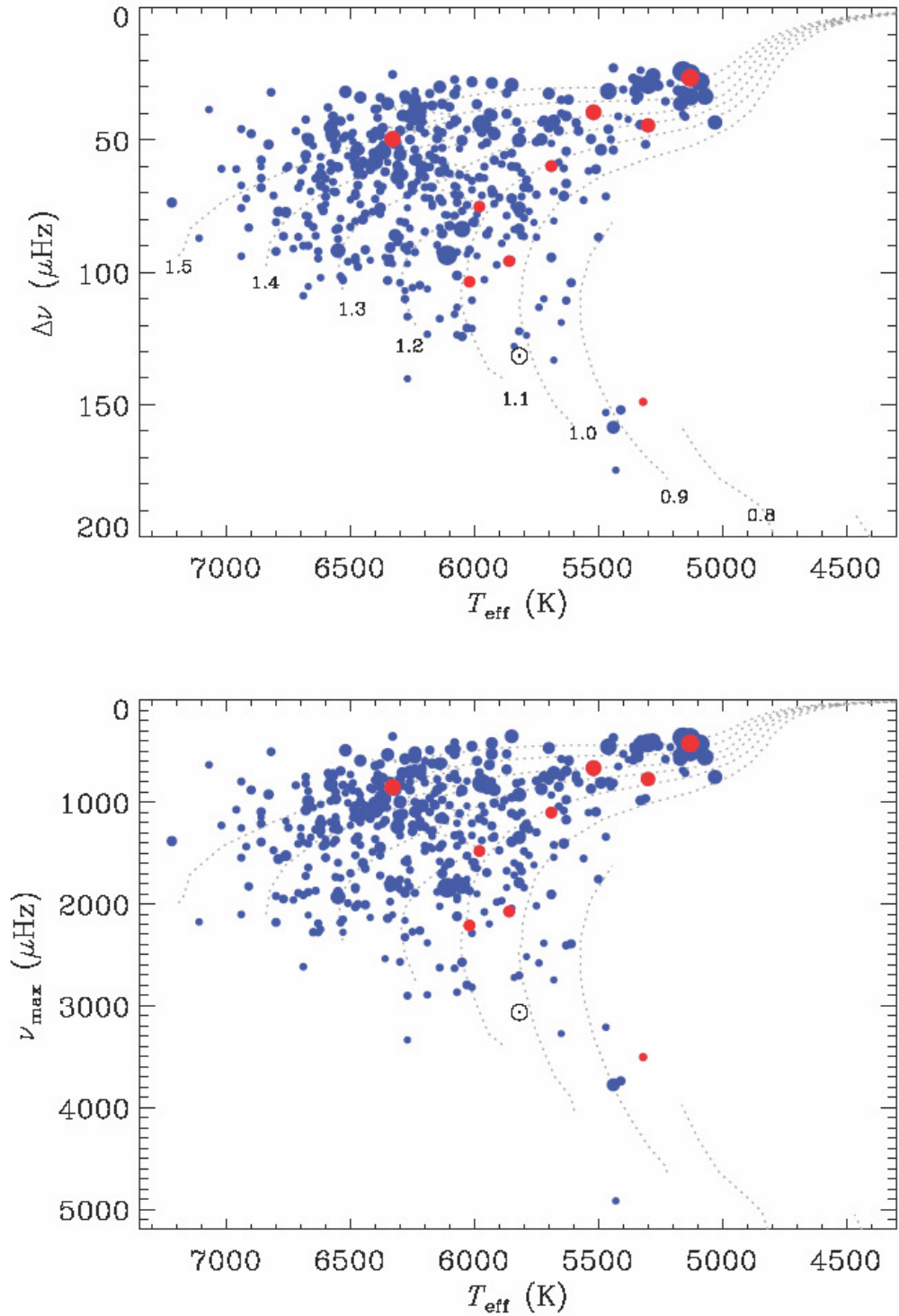

Figure S1: Large frequency separations, $\Delta\nu$, and frequencies of maximum oscillations power, $\nu_{max}$, plotted as a function of effective temperature, $T_{eff}$. (See also Fig. 2 in main article.)

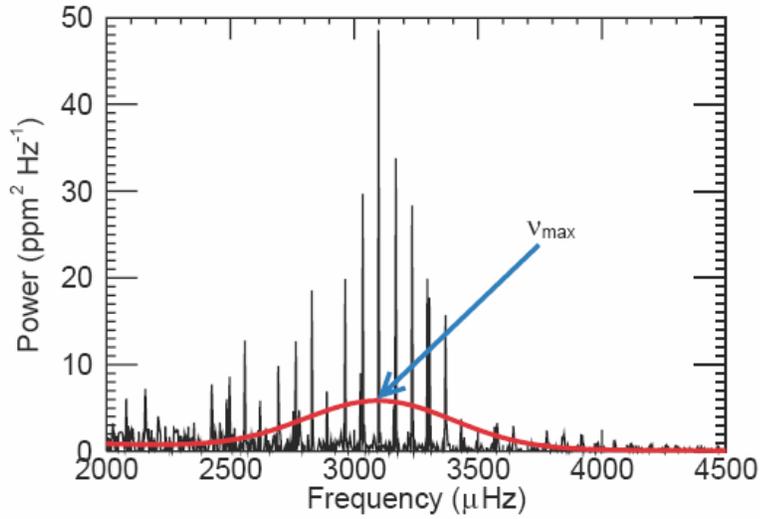

Figure S2: Frequency spectrum of low-degree oscillations shown by the Sun (in Sun-as-a-star photometry data from the VIRGO/SPM instrument on board the ESA/NASA SoHo spacecraft). The red line follows the Gaussian-like power envelope of the observed oscillations, with the frequency of maximum power marked by $\nu_{max}$.

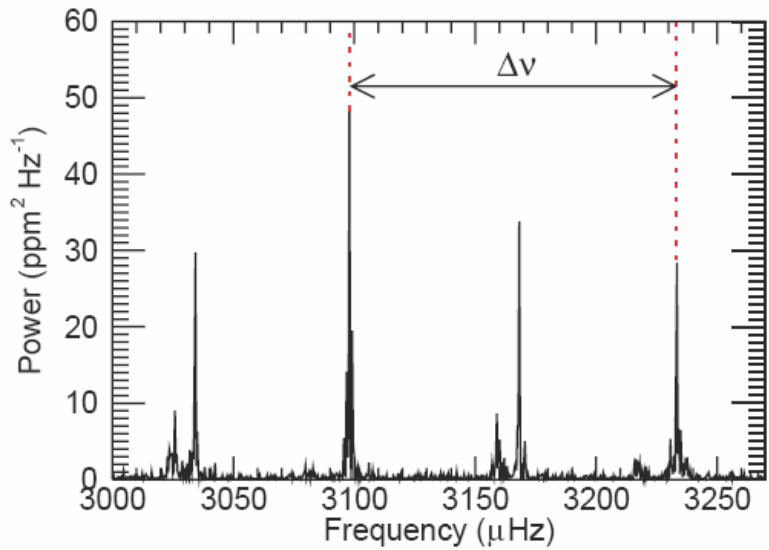

Figure S3: Zoom on the central frequency region of the oscillations spectrum plotted in Fig. S2, to show the large frequency separation, $\Delta\nu$.